\documentclass[twocolumn,prl,floats]{revtex4}
\usepackage{graphicx}
\usepackage{amssymb}

\newcommand{\vect}[1]{{\mbox{\boldmath $#1$}}}

\begin{document}

\title{Perturbation of magnetostatic modes observed by FMRFM}

\author{R. Urban$^1$, A. Putilin$^1$, P.E. Wigen$^{1,2}$, S.-H. Liou$^3$, M.C. Cross$^1$, P.C. Hammel$^2$, and M.L. Roukes$^1$}
\affiliation{$^1$Condensed Matter Physics, California Institute
of Technology, MC 114-36, Pasadena, CA 91125, USA \\
$^2$Department of Physics, Ohio State University, Columbus, OH 43210, USA \\
$^3$Department of Physics and Astronomy, University of Nebraska,
Lincoln, NE 68588, USA}

\date{\today}


\begin{abstract}
Magnetostatic modes of Yttrium Iron Garnet (YIG) films are
investigated by ferromagnetic resonance force microscopy (FMRFM).
A thin film ``probe'' magnet at the tip of a compliant cantilever
introduces a local inhomogeneity in the internal field of the YIG
sample.  This influences the shape of the sample's magnetostatic
modes, thereby measurably perturbing the strength of the force
coupled to the cantilever.  We present a theoretical model that
explains these observations; it shows that tip-induced variation
of the internal field creates either a local ``potential barrier''
or ``potential well'' for the magnetostatic waves. The data and
model together indicate that local magnetic imaging of
ferromagnets is possible, even in the presence of long-range spin
coupling, through the induction of localized magnetostatic modes
predicted to arise from sufficiently strong tip fields.
\end{abstract}

\maketitle

%
In the past decade, the possibility of using magnetic degrees of
freedom in electronic devices has attracted the attention of the
semiconductor industry. There are numerous proposals to
incorporate spin degrees of freedom into electronics creating
spin-electronic or spintronic devices \cite{Heinrich-CJP00}. These
devices require a thorough understanding of the magnetization
dynamics on the submicron scale. Magnetic resonance force
microscopy (MRFM) has proven itself to be a three dimensional
nondestructive imaging technique that can be applied to electron
\cite{Rugar-Nature92,Hammel-JLTP95,Rugar-Nature04} and nuclear
\cite{Rugar-Science94} resonance experiments. In MRFM a
micromagnet (tip magnet) is placed on a compliant cantilever. The
time variation of the amplitude of the precessing spins will
produce a force on the cantilever via the gradient field of the
tip magnet. In addition the tip magnet produces a perturbation in
the magnetic field in the sample producing a small bowl shaped
region in which the spins are at resonance, the resonance slice.
This resonance slice can be highly localized (on the order of the
size of the tip magnet) thus allowing subsurface detection of
spins with submicron resolution. A recent experiment reported the
detection of a single spin in resonance \cite{Rugar-Nature04}.

Ferromagnetic Resonance Force Microscopy (FMRFM)
\cite{Zhang-APL96,Loehndorf-APL00,Charbois-JAP02} is a variation
of MRFM that enables the three dimensional, high sensitivity and
high resolution qualities of MRFM to be applied in the
characterization of the dynamic magnetic properties of
ferromagnetic thin films and multilayers at the micron scale. The
tip field penetrates magnetic and nonmagnetic materials allowing
one to investigate buried structures and interfaces. Recent
publications have shown the ability of FMRFM to determine the
nature and magnitude of terms contributing to the internal field
\cite{Loehndorf-APL00}, dispersion relations of magnetostatic
modes \cite{Zhang-APL96,Charbois-JAP02,Midzor-JAP00} and
relaxation processes in micron size samples \cite{Klein-PRB03}.
Ferromagnetically coupled systems pose unique challenges for
magnetic resonance imaging due to the strong coupling between the
spins. The resulting magnetostatic/exchange resonance modes
involve spins occupying the entire sample not just those within
the resonance slice \cite{Midzor-JAP00}. Therefore, local imaging
capability of MRFM is lost in the case of ferromagnetic samples.

In this Letter we report the first direct observation of the
variation of the magnetostatic mode amplitudes due to the local
perturbation of the internal field. The localized field is
produced by the sharp magnetic tip when it is scanned across the
surface of Yttrium Iron Garnet (YIG) samples. The experimental
results are in excellent agreement with theoretical predictions.
In addition, based on our theoretical model, we propose that it is
possible to generate a highly localized mode and recover, in part,
the advantages of local imaging capabilities of MRFM. These
localized magnetic modes will provide an exciting opportunity to
investigate static and dynamic magnetic properties of micron-size
defects (imperfections) embedded in ferromagnetic samples without
the interference of edge effects.

%
The experiments were performed using an ambient {FMRFM} system in
the perpendicular geometry \cite{Midzor-JAP00}. The sample is
attached to a microstrip resonator having a resonance frequency of
7.4 GHz. The silicon cantilever's resonance frequency, the spring
constants, and the ambient Q-factor are 17.8 kHz, 0.4 N/m, and 90,
respectively. The cantilever displacement is monitored by 850 nm
fiber optic interferometer. The cantilever tip was coated with a
250 nm CoPt film ($M_s = 800$ emu/cm$^3$) and annealed in an
external magnetic field of 80 kOe oriented along the tip axis of
the cantilever. This resulted in a coercive field of 10 kOe
\cite{Signoretti-JMMM04}. The magnetic tip generates a localized
tip field, $h_{\rm tip}$, which can be aligned either parallel
($h_{\rm tip} > 0$) or antiparallel ($h_{\rm tip} < 0$) to the
external dc field, $H_{\rm ext}$. Based on a model of the magnetic
tip, the strength of the tip field and the tip field gradient at
the tip-sample separation $z = 1$ $\mu$m were estimated to be 80 G
and 50 G/$\mu$m for $h_{\rm tip} > 0$.

\begin{figure}[t!] %
    \begin{center}
        \includegraphics[scale=0.65]{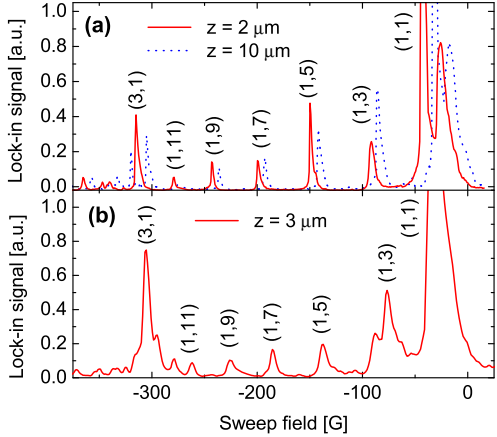} %
        \vskip-6.5mm %
    \end{center}
    \caption{Typical FMRFM spectra measured in the center ($y = 0$ $\mu$m) of the
    $20 \times 80 \, \mu$m$^2$ sample for (a) $h_{\rm tip} > 0$ and (b) $h_{\rm tip} < 0$,
    respectively. The dashed and the solid lines correspond to
    the tip-sample separation of 10 $\mu$m and 2 $\mu$m,
    respectively. Zero sweep field corresponds to the resonance
    field of the homogeneous mode $\omega / \gamma + 4 \pi M_s =
    4.4$ kG.
    } %
    \label{fig1} %
\end{figure}

In the following section we present experimental data that confirm
the importance of local probe-induced phenomena. Typical measured
spectra obtained on the $20 \times 80$ $\mu$m$^2$ YIG sample are
shown in Fig. \ref{fig1}. The magnetostatic modes are labelled by
$(n_x,n_y)$, where $n_x$ and $n_y$ represent the number of half
wavelengths along the sample width and length, respectively. In
the presence of a uniform bias field and a homogeneous driving rf
field only modes with odd values of $n_x$ and $n_y$ have a net
dipole moment that will couple to the rf field
\cite{Wigen-Hilleb04}. However, if the symmetry of the magnetic
field is broken by the presence of the localized tip field, the
magnetostatic modes having $n_x$ and/or $n_y$ even can have a net
dipole moment and be excited. We now discuss the cases of parallel
and antiparallel tip orientations.

\vect{h_{\rm tip} > 0}: To investigate the effects of this
perturbation quantitatively, we first discuss the case when the
magnetic field of the tip magnet adds to the bias magnetic field.
For a weak tip field ($h_{\rm tip} \approx 0$) applied at the
center of the sample the mode amplitudes are observed to decrease
monotonically with increasing $n_x$ and $n_y$. Such a spectrum is
shown by the dashed line in Fig. \ref{fig1}a for $z = 10$ $\mu$m.
At that distance the magnetic field generated by the tip magnet is
on the order of 2 G. As the tip approaches the sample surface, the
magnetic tip field increases. This results in a dramatic change of
the measured spectra as illustrated by the solid line in Fig.
\ref{fig1}a. There are two main features to be noted: (i) the
spectrum is shifted to a lower resonance field by $(8 \pm 1)$ G
and (ii) the amplitude of modes (1,5) and (1,9) are enhanced
relative to modes (1,3), (1,7), and (1,11). The resonance field
shift is due to the spatially-averaged tip field which adds to the
external field. Notice, that the principal mode ($n_x = n_y = 1$)
is also split. We attribute the mode on the high field side to be
a localized surface mode \cite{Wigen-TSF84}.

\begin{figure}[t!] %
    \begin{center}
        \includegraphics[bb=25 20 700 240,scale=0.42]{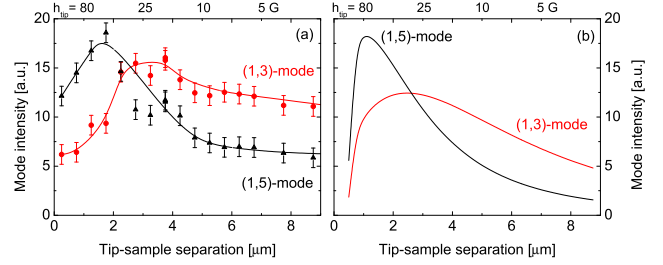} %
        \vskip-6.5mm %
    \end{center}
    \caption{(a) The mode intensity of the (1,3) and (1,5) mode as a
    function of the tip-sample separation. The tip was placed
    above the middle of the $20 \times 80$ $\mu$m$^2$ sample. The
    solid lines are a spline fit to data to guide the eye.
    (b) The theoretical prediction using a cylindrical magnetic tip.
    } %
    \label{fig2} %
\end{figure}

To demonstrate the subtle details of the effect of the increasing
tip field, the force intensity for modes (1,3) and (1,5) is
plotted in Fig. \ref{fig2} as a function of the tip-sample
separation. As the tip field gradient increases with decreasing
distance the FMRFM signal increases. However, the intensity of the
(1,3) mode decreases dramatically for $z < 3$ $\mu$m and at
$z=2.5$ $\mu$m its intensity becomes even lower than the intensity
of the (1,5) mode. For $z < 2$ $\mu$m the intensity of the (1,5)
mode decreases as well. These effects can be explained by the
model proposed below, see Fig. \ref{fig2}b.
\begin{table}[b!]
  \centering
\begin{tabular}{cccc}
        \hline \hline
    & I(1,3)/I(1,5) \qquad & I(1,5)/I(1,7) \qquad & I(1,7)/I(1,9) \\ \hline
    $h_{\rm tip} > 0$ & $0.6 \pm 0.1$ & $2.9 \pm 0.1$ & $1.0 \pm 0.2$ \\
    $h_{\rm tip} \approx 0$ & $1.6 \pm 0.1$ & $1.9 \pm 0.1$ & $1.5 \pm 0.2$ \\
    $h_{\rm tip} < 0$ & $2.4 \pm 0.2$ & $1.1 \pm 0.3$ & $1.9 \pm 0.4$ \\
    \hline \hline
\end{tabular}
  \caption{The intensity ratios for modes (1,3), (1,5), (1,7), and
  (1,9) in the presence of a strong positive tip field ($h_{\rm tip} >
  0$), a weak tip field ($h_{\rm tip} \approx 0$), and a negative
  tip field ($h_{\rm tip} < 0$). Corresponding spectra are plotted
  in Fig. \ref{fig1}.
  } %
  \label{ratio} %
\end{table}

\vect{h_{\rm tip} < 0}: By reversing the orientation of the
applied magnetic field without reversing $h_{\rm tip}$, the
magnetic field of the tip subtracts from the homogeneous bias
field. For a tip-sample separation below 2 $\mu$m, the modes
(1,3), (1,7), and (1,11) are now stronger compared to modes (1,5)
and (1,9), see Fig. \ref{fig1}b. This is exactly opposite to what
is observed for $h_{\rm tip} > 0$. Table \ref{ratio} summarizes
the intensity ratios of the subsequent resonant modes when the tip
field is weak and the magnetostatic modes are practically
unperturbed, $h_{\rm tip} \approx 0$ line and for strong tip
fields, $h_{\rm tip} > 0$ and $h_{\rm tip} < 0$. There are two
notable features: (i) the values in the first row are
lower/higher/lower than values for $h_{\rm tip} \approx 0$ while
the values in the third row are higher/lower/higher than the
values for $h_{\rm tip} \approx 0$. This reinforces the
conclusions made earlier that for $h_{\rm tip}
> 0$, the modes (1,5) and (1,9) are enhanced, while for $h_{\rm
tip} < 0$ the modes (1,3) and (1,7) are enhanced. (ii) The first
and third columns are ascending while the second column is
descending.

\begin{figure}[t!] %
    \begin{center}
        \includegraphics[scale=0.6]{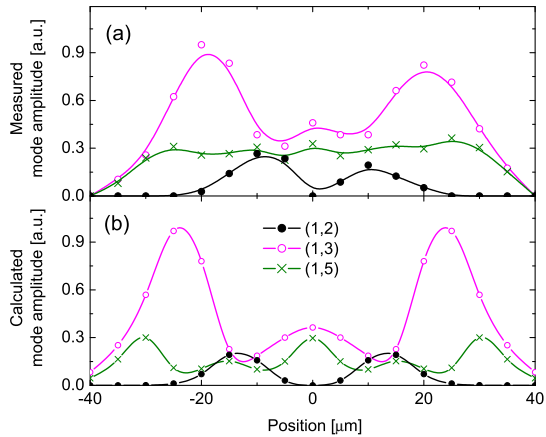} %
        \vskip-6.5mm %
    \end{center}
    \caption{The mode intensity for the (1,2), (1,3), and (1,5) modes as the
    tip is scanned along the long axis of the $20
    \times 80 \, \mu$m$^2$ sample. The external dc field and
    the probe field are parallel. The tip-sample separation
    $z \approx 3$ $\mu$m. (a) Measured response. (b) Calculated
    amplitude.
    } %
    \label{fig3} %
\end{figure}

The influence of the perturbation field of the probe magnet is
further demonstrated in lateral scans taken along the long axis of
the sample at $z \approx 3$ $\mu$m. Fig. \ref{fig3}a shows the
experimental data of the force amplitudes for the ($1,n_y$) modes
as a function of position. Three interesting features are
observed: (i) As the tip is moved from the center of the sample,
the  field of the tip magnet has broken the even symmetry of the
internal field. As a result the ``hidden'' (1,2) mode is excited.
(ii) For the (1,3) mode, the intensity at $y = 0$ $\mu$m is
strongly suppressed while at the position of the next maxima in
the magnetostatic mode at $y = \pm 26$ $\mu$m the intensity is
enhanced. (iii) Mode (1,5) shows very little variation as a
function of position of the tip. Therefore, the lateral resolution
can be estimated to be approximately 20 $\mu$m. This is in good
agreement with the theoretical predictions (Fig. \ref{fig3}b).
%

These unusual characteristics are understood within a theoretical
model which is based on a linearized Landau-Lifshitz-Gilbert (LLG)
equation of motion that include a spatially dependent tip field
and Maxwell's equations. The external dc field is parallel to the
sample normal and the driving rf field, $h_{\rm rf}$, lies in the
plane of the sample and is considered to be homogeneous over the
entire sample. The cantilever tip field, $h_{\rm tip}(\vect{r})$,
is considered to be either parallel, $h_{\rm tip}
> 0$, or antiparallel, $h_{\rm tip} < 0$, to the external field
$\vect{H}_{\rm ext}$.

The linearized LLG equation of motion and Maxwell's equations
neglecting displacement currents can be written as (in cgs units)
\begin{eqnarray}
\frac{\partial \vect{M}}{\partial t} = - \gamma \left[ \vect{M}
\times \vect{H}_{\rm eff} \right] + \frac{\alpha}{M_s} \left[
\vect{M}
\times \frac{\partial \vect{M}}{\partial t} \right] \; , \label{eqn:LLG}\\
\nabla \times \vect{H} = 0 \; , \qquad \nabla \cdot (\vect{H} +
4\pi\vect{M}) = 0 \, , \label{eqn:Maxwell}
\end{eqnarray}
where $\gamma$ ($= 1.84 \times 10^7$ Oe$^{-1}$s$^{-1}$) is the
absolute value of gyromagnetic ratio, $M_s$ ($= 140$ erg/cm$^3$)
is the saturation magnetization, and $\alpha$ ($= 0.0045$) is the
dimensionless Gilbert damping parameter. The internal effective
field, $\vect{H}_{\rm eff}$, is given by
\begin{eqnarray}
\vect{H}_{\rm eff} = \left( h_x + h_{\rm rf}, h_y, H_{\rm
ext}-4\pi M_s + h_{\rm tip}\left( \vect{r} \right) \right) \, .
\end{eqnarray}
Eq. (\ref{eqn:LLG}) is solved in a small angle approximation;
$|\vect{m}| \ll M_s$, where \vect{m} represents the transverse rf
component of the magnetization vector. The boundary conditions are
assumed to be
\begin{eqnarray} \label{eqn:bc}
\vect{m}|_{x=\pm L_x/2} = \vect{m}|_{y=\pm L_y/2} = 0 \, .
\end{eqnarray}
%

The unique feature in the model is treating the tip field as a
non-local perturbation and determining its effect on the
magnetostatic modes. Using Eq. \ref{eqn:Maxwell}, the rf magnetic
field $\vect{h}$ is expressed in terms of the magnetization
$\vect{m}(\vect{r})$ as a linear functional
$\vect{h}[\vect{m}(\vect{r})]$ resulting in the non-local
relationship
\begin{eqnarray} \label{eqn:master}
\hat{K} \vect{m}(\vect{r}) + h_0(\vect{r}) \vect{m}(\vect{r}) =
\Delta \Omega \vect{m}(\vect{r}) + h_{\rm rf} \, , \\
\label{eq:energy} \Delta \Omega = \frac{1}{4\pi M_s} \left(
\frac{\omega}{\gamma} -
H_{\rm ext} + 4 \pi M_s \right) \, , \\
\label{eq:tip_field} h_0(\vect{r}) = \frac{h_{\rm
tip}(\vect{r})}{4\pi M_s} \, .
\end{eqnarray}
The operator $\hat{K}$ represents a linear integral transformation
defined by
\begin{eqnarray} \label{eqn:SR}
\hat{K} f_{\vect{k}}(\vect{r}) = \omega_{\vect{k}}
f_{\vect{k}}(\vect{r}) \, ,
\end{eqnarray}
where $f_{\vect{k}}$ represents the unperturbed magnetostatic
modes given by
\begin{eqnarray}
f_{\vect{k}}(x,y) = \sin{\left\{k_x \left( x+ L_x/2 \right)
\right\}} \, \sin{\left\{k_y \left( y+ L_y/2 \right) \right\} }
\end{eqnarray}
and $\omega_{\vect{k}}$ is their dispersion relation given by the
Damon-Eshbach theory \cite{Eshbach-PR60}. The wave-vectors
$\vect{k}$ are chosen to satisfy the ``pinned'' boundary
conditions, Eq. \ref{eqn:bc}. The new basis of the magnetostatic
modes ${m}_n(\vect{r})$ can be found by solving the homogeneous
equations
\begin{eqnarray} \label{eqn:Shrodinger}
\left[ \hat{K} + h_0(\vect{r}) \right] m_n(\vect{r}) = \Delta
\Omega_n m_n(\vect{r}) \, .
\end{eqnarray}

The FMRFM signal is proportional to the force acting on the
cantilever which can be written as
\begin{eqnarray}
F_{\rm FMRFM} \approx d \int_{A} M_z(\vect{r}) \cdot \nabla_z
h_{\rm tip}(\vect{r}) \, {\rm d}x {\rm d}y \, ,
\end{eqnarray}
where $M_z(\vect{r}) = M_s - |\vect{m}|^2/2 M_s$ is the
longitudinal component of magnetization, $d$ is the thickness of
the film and $A$ represents the area of the sample. A detailed
discussion of the theory will be presented elsewhere
\cite{Putilin-}.

Eq. \ref{eqn:Shrodinger} offers an interesting interpretation; It
is similar to the time-independent Schr{\" o}dinger equation with
the kinetic energy-like term given by $\omega_{\vect{k}}$ in Eq.
\ref{eqn:SR}. The normalized tip field $h_0(\vect{r})$ in Eq.
\ref{eq:tip_field} plays the role of a potential energy. When the
tip field is parallel to the external field, $h_0(\vect{r}) > 0$,
it creates a {\it potential barrier} for the magnetostatic modes,
while $h_0(\vect{r}) < 0$ yields a {\it potential well}.
\begin{figure}[t!] %
    \begin{center}
    \vskip-3mm
        \includegraphics[bb=0 0 300 350,scale=0.51]{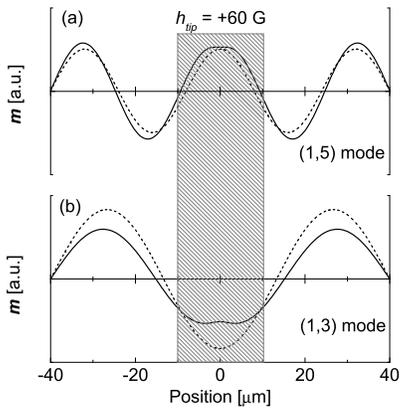} %
        \vskip-6.5mm %
    \end{center}
    \caption{The calculated magnetostatic modes
    of the $20 \times 80$ $\mu$m$^2$ sample with the tip located at the
    center of the film. The unperturbed (dashed) and
    perturbed (solid) modes are plotted for a potential barrier $h_{\rm tip} = + 60$ G.
    Notice, that the wavelength of the mode is increased in the
    perturbed region (shaded). (a) For the (1,5) mode this region
    increases the net dipole moment and therefore increases the FMRFM signal.
    (b) For (1,3) mode this region
    decreases the net dipole moment and therefore reduces the FMRFM signal.
    } %
    \label{fig4} %
\end{figure}
This analogy provides qualitative insight into how the
magnetostatic modes are modified by the tip magnetic field.

\vect{h_{\rm tip} > 0}: The effect of the perturbation field when
the tip field is parallel to the bias field is to increase the
wavelength of the magnetostatic mode in the region near the tip.
Fig. \ref{fig4} shows this effect when the tip is located at the
center of the film. If \vect{m} is positive (i.e., when \vect{m}
is parallel to the net dipole moment of the mode) in the region
near the tip as in the case of mode (1,5) shown in Fig.
\ref{fig4}a, the net dipole moment of the mode will be increased
and therefore the FMRFM signal is enhanced. On the other hand, if
\vect{m} is negative (i.e., when \vect{m} is antiparallel to the
net dipole moment of the mode) in the region of the tip field as
shown for mode (1,3) in Fig. \ref{fig4}b, the net dipole moment
will be decreased and therefore the FMRFM signal for that mode is
suppressed.

As the tip approaches the surface of the sample the tip field
gradient increases and therefore the FMRFM signal increases, see
Fig. \ref{fig2}. However, when the field separation of the
magnetostatic mode $4 \pi M_s \Delta \Omega$ (see Eq.
\ref{eq:energy}) is smaller than the tip field $h_{\rm tip}$, the
mode becomes evanescent. This decreases the amplitude of the
magnetostatic wave in the vicinity of the tip magnet resulting in
a reduced FMRFM signal  even though the gradient of the field is
increased. This is demonstrated in Fig. \ref{fig2}b.

When the tip magnet is scanned across the sample, the transverse
component of the magnetization \vect{m} changes sign at the next
lobe position; At $y = 0$ $\mu$m, the transverse component of
\vect{m} for the (1,3) mode is negative and therefore the
intensity of this mode is strongly suppressed. On the other hand,
at $y = \pm 26$ $\mu$m, \vect{m} is positive and, as observed in
Fig. 4 this mode is strongly enhanced.

\vect{h_{\rm tip} < 0}: In the case of the potential well the
results are exactly the opposite to those for a potential barrier
($h_{\rm tip} > 0$). If \vect{m} is negative in the region of the
tip magnet as is the case for mode (1,3), the wavelength of the
wave in the region of the perturbing tip field is decreased and
the coupling of the mode to the rf field is increased.
Consequently the FMRFM intensity of the mode is enhanced. On the
other hand, if \vect{m} is positive in the region of the tip field
as for mode (1,5), the net dipole moment will be decreased and
therefore the FMRFM signal for that mode is suppressed.

Extending the Schr{\" o}dinger equation analogy to the case of a
deep narrow potential well, a highly localized magnetic mode can
be excited \cite{Putilin-}. The properties of this localized mode
would be independent of the sample size and therefore of any edge
imperfections that might be introduced in the sample preparation.
Scanning the local mode about the sample would allow one to
extract variations in the local magnetic properties of the
ferromagnetic samples. This makes FMRFM a unique experimental
method allowing the investigation of local magnetic properties of
the ferromagnetic samples independent on the sample dimensions,
shape and/or defects at the sample edge. Incorporation of advanced
nanomagnetic probe tips should enable next-generation FMRFM
systems capable of imaging local magnetic properties with
sub-micron lateral resolution.

%
In conclusion, it has been demonstrated that the magnetic field of
the tip magnet introduces a local inhomogeneity in the internal
field of a $20 \times 80 \, \mu$m$^2$ YIG film. This influences
the shape of the samples magnetostatic modes, thereby measurably
perturbing the strength of the force coupling to the cantilever.
Using a hard magnetic coating on the tip, we are able to
investigate the magnetostatic modes when the tip field is parallel
($h_{\rm tip} > 0$) and antiparallel ($h_{\rm tip} < 0$) to the
external dc field. The condition $h_{\rm tip} > 0$ can be
visualized as a potential barrier in the Schr{\" o}dinger-like
equation analogy, and modes (1,5) and (1,9) are enhanced while
modes (1,3) and (1,7) were suppressed. In contrast with this is
the case for $h_{\rm tip} < 0$ corresponding to a potential well
in the Schr{\" o}dinger-like equation analogy. Here the observed
effect is exactly the opposite to the case for $h_{\rm tip} > 0$.
These results are in excellent agreement with the proposed theory
indicating, that an inhomogeneity in the internal field can also
be successfully included into FMRFM experiments. Furthermore,
highly localized magnetostatic modes are predicted which are
suitable for {\bf local imaging} of ferromagnetic samples.

%
We thank to Melissa M. Midzor for assistance and discussions. We
gratefully acknowledge financial support from ONR MURI under grant
DAAD 19-01-0541. P.E.W. was supported by R.J. Yeh fund during the
course of this research.


\end{document}